\documentclass[twocolumn,aps,prd]{revtex4}%
\usepackage{graphicx}
\usepackage{epsf}
\usepackage{epsfig}
\usepackage{bm}
\usepackage{amsmath}
\usepackage{dcolumn} 
\usepackage{amsfonts}
\usepackage{amssymb}%
\providecommand{\U}[1]{\protect\rule{.1in}{.1in}}
\newcommand{\be}{\begin{equation}}
\newcommand{\ee}{\end{equation}}

\newcommand{\mincir}{\raise
-3.truept\hbox{\rlap{\hbox{$\sim$}}\raise4.truept\hbox{$<$}\ }}
\newcommand{\magcir}{\raise
-3.truept\hbox{\rlap{\hbox{$\sim$}}\raise4.truept\hbox{$>$}\ }}

\begin{document}

\title{Is $\Lambda$CDM an effective CCDM cosmology?}

\author{J. A. S. Lima\footnote{limajas@astro.iag.usp.br}}

\affiliation{Departamento de Astronomia, Universidade de S\~ao Paulo, Rua
do Mat\~ao 1226, 05508-900, S\~ao Paulo, SP, Brazil}

\author{R. C. Santos\footnote{cliviars@gmail.com}}

\affiliation{Departamento de Ci\^encias Exatas e da Terra, Universidade Federal de S\~ao Paulo\\ 09972-270  Diadema, SP, Brasil} 

\author{J. V. Cunha\footnote{jvcunha@ufpa.br}}

\affiliation{Faculdade de F\'{i}sica, Universidade Federal do Par\'a, 66075-110, Bel\'em, PA, Brazil} 

\begin{abstract} 

\noindent We show that a cosmology driven by gravitationally induced particle production of all non-relativistic species existing in the present Universe mimics exactly the observed flat accelerating  $\Lambda$CDM cosmology with just one dynamical free parameter. {This kind of scenario includes the creation cold dark matter (CCDM) model [Lima, Jesus \& Oliveira, JCAP 011(2010)027] as a particular case} and also provides a natural reduction of the dark sector since the vacuum component is not needed to accelerate the Universe. The new cosmic scenario is equivalent to $\Lambda$CDM  both at the background and perturbative levels and the associated creation process is also in agreement with the universality of the gravitational interaction and equivalence principle. Implicitly, it also suggests that the present day astronomical observations cannot be considered the ultimate proof of cosmic vacuum effects in the evolved Universe because $\Lambda$CDM may be only an effective cosmology. 

  
\end{abstract}
\pacs{98.80.-k, 95.35.+d, 95.36.+x}
\keywords{Cosmology; dark energy}\maketitle

\hyphenation{tho-rou-ghly in-te-gra-ting e-vol-ving con-si-de-ring
ta-king me-tho-do-lo-gy fi-gu-re}

\noindent {\bf 1. Introduction.\,} The present cosmic concordance model ($\Lambda$CDM cosmology) is plagued with two profound mysteries  which are  challenging our present understanding of fundamental  physics: the cosmological constant and coincidence problems (CCP and CP). The former is directly related to the huge discrepancy between the vacuum energy density determined from astronomical observations and the value ranging from 50-120 orders of magnitude larger, as expected from quantum field theory with a convenient high-energy cut-off \cite{Weinberg89}. The latter one (CP) arises because the constant vacuum and decreasing matter energy densities are now finely tuned, but their ratio was incredibly small in the distant past\cite{Coincidence}. 

Dozens of models based on disparate theoretical frameworks were proposed in the last decade trying  to circumvent the aforementioned  cosmological puzzles. Some attempts  includes cosmologies based on modified gravity theories, quintessence dark energy models, dynamical vacuum models, and the influence of inhomogeneities during the nonlinear stage of the structure formation process \cite{Lambda}. {For instance, although also in agreement with the present observations, many dynamical $\phi CDM$ dark energy models originally proposed as alternatives to $\Lambda$CDM  do not solve (or alliviate)  both the CCP and CP problems. New fine-tunings are usually introduced in the form of small masses and couplings of the new scalar field. Possible exceptions are scenarios where the de Sitter spacetime is a final attractor solution thereby contributing to alleviating the fine-tunning problem \cite{Peebles88}.}  

Although extensively investigated in the literature, dark energy models seem to be less competitive than the one free parameter cosmic concordance  model \cite{Reviews}. {In a point of fact, studies applying direct comparison methods, like the Bayesian information criterion (BIC), shown that the $\Lambda$CDM model is the most favored accelerating cosmology at light of the present day observations, a result to some extent related to the so-called Occam's razor \cite{Occam}.}  

In this context, one may assume on phenomenological grounds that $\Lambda \equiv 0$. However, once this solution is adopted a new question arises: {\it what kind of one-parametric cosmology will replace the standard flat $\Lambda$CDM concordance model?} A positive answer is required here because if a new cosmology is called for it must looks very much like $\Lambda$CDM, the model currently preferred by all available astronomical observations (SNe Ia, CMB, BAO, Large Scale Structure, Clusters, H(z), etc). 

It is also widely known that a general relativistic (GR) model dominated only by non-relativistic components (CDM and baryons) requires an extra mechanism in order to account for the present accelerating stage. The adopted mechanism must provide a new cosmic concordance cosmology without vacuum contribution and whether possible none kind of coincidence problem.  

On the other hand, a great deal of attention has recently been paid for accelerating cosmologies driven by ``adiabatic" particle production  where matter and entropy are generated but the specific entropy (per particle) remains constant\cite{LJO2010,BL2010,BL2011,MP13,Komatsu2014,Pan,Waga2014,Waga2014a}.  
The macroscopic foundation of the negative pressure accompanying particle creation  is self-consistently  derived using relativistic non-equilibrium thermodynamics \cite{Prigogine89,LCW92,CLW92}.  {The second law of thermodynamics determines how an irreversible process of quantum origin can be incorporated into the classical Einstein field equations. The negative creation pressure acts like a second viscosity stress, a mechanism suggested much earlier by Zeldovich [15]. However, as discussed in detail by Lima and Germano \cite{LG92}, it cannot describe particle production because the corresponding thermal evolution is fully different even when the same dynamics is fixed. The relativistic kinetic theory describing the gravitationally induced particle production process has also been discussed in the recent literature\cite{LI2014,IB2015}. The proposed equation generalizes the  standard mass-shell  of the standard Boltzmann equation without gravitational particle production, as ordinarily presented in textbooks. This  kinetic approach  gives rise to the possibility of a new formalism  inspired on the extended Boltzmann equation where back reaction effects and constraints from the second law of thermodynamics might be naturally incorporated in the spacetime description.}  In principle, such requirements are beyond the standard quantum field theory in curved spacetimes based on the ideas of adiabatic vacuum and amplitudes (for particle production) calculated through the Bogoliubov mode mixing technique\cite{BD89, MS08, PT09} for a test field. {However, its
precise determination requires an acceptable non-equilibrium theory for gravitational induced particle production using finite-temperature
quantum field theory in curved space-times. The lack of such a theory suggests naturally a phenomenological approach by incorporating back-reaction in the context of general relativity theory.}

Current cosmological scenarios with creation are based only on gravitationally induced production of CDM particles, and, as such, are usually dubbed CCDM cosmologies. Although realistic in several aspects (see quoted works\cite{LJO2010,BL2010,BL2011,MP13,Komatsu2014,Pan,Waga2014,Waga2014a}),  such  models are  inherently  incomplete from a theoretical viewpoint. Indeed, the gravitational particle production of only a given component (CDM) challenges the universality of the gravitational interaction, unless some hidden (unknown) symmetry is forbidding the multi-particle creation by the gravitational field. 

In this {\it Letter}, we discuss how gravitationally induced particle production of an arbitrary number of non-relativistic cosmic components (including baryons)  affect the original CCDM cosmology\cite{LJO2010,LI2014} (see also \cite{Zimd2001}). As we shall see, beyond to solve the mentioned incompleteness it is found that the  $\Lambda$CDM model can be replaced by a new cosmology  based on the gravitational particle production of all self-gravitating fields existing in the Universe. Amazingly, the resulting cosmology mimics the flat $\Lambda$CDM model both at background and perturbative levels (linear and nonlinear) with just one free parameter describing the effective creation rate of all existing non-relativistic components.  
\vspace{0.2cm}

\noindent {\bf 2. Extended CCDM  cosmology}

\vspace{0.2cm}
Let us now consider that the Universe is described by a flat Friedmann-Robertson-Walker (FRW) geometry:

\begin{equation}
 ds^2 = dt^2 - a^{2}(t)({dx^2} +  dy^{2}  +  dz^{2}),
\end{equation}
where $a(t)$ is the scale factor.  In such a background, the Einstein equations (without $\Lambda$) for a N-component fluid mixture  endowed with gravitationally induced ``adiabatic" particle production reads: 

\begin{eqnarray}
&&8\pi G\rho_{\rm T}\equiv 8\pi G\sum_{i=1}^{N} \rho_i =3\frac{{\dot a}^{2}}{a^2}\,,\label{EE1}\\
&&8\pi G p_{\rm T} \equiv 8\pi G \sum_{i=1}^{N} (p_i + P_{ci})=-2\frac{\ddot a}{a}-\frac{{\dot a}^{2}}{a^2}\label{EE2}\,,
\end{eqnarray}
where an over-dot means time derivative with respect to the cosmic time, $\rho_i$ (i=1,2,..N), is the energy density of an arbitrary created component. The quantities, $p_i$,   $P_{ci}$, denote the thermodynamic (kinetic) and creation pressures, respectively. The latter is given by \cite{Prigogine89,LCW92}: 
\begin{equation}\label{pressaotermo}
 P_{ci}=-(\rho_i + p_i)\frac{\Gamma_i}{3H}= - (1+ \epsilon_i) \frac{\rho_i \Gamma_i}{3H}, 
\end{equation}
where $\Gamma_i$  is the creation rate of each component, $H={\dot a}/a$ is the Hubble parameter, and, for the second equality above, we have adopted  the usual EoS  
\begin{equation}\label{EoS}
 p_i = \epsilon_i \rho_i\,,\,\,\,(\epsilon_i=const \geq 0).
\end{equation}
Hence, in this kind of scenario without dark energy, the standard kinetic pressure, $p_i$, is always positive while the present accelerating regime is provoked by the negative creation pressure, $P_{ci}$.

The ratio $\Gamma_i/H$ in Eq.(\ref{pressaotermo}) quantifies the efficiency of the gravitational creation process. In particular, if $\Gamma_i << H$ it can be safely neglected. The gravitational production process as described here occurs in such a way that the specific 
entropy (per particle) of each component is constant (``adiabatic"' gravitational particle production), that is, $\sigma_i = S_i/N_i=constant$, where $S_i=s_ia^{3}$ and $N_i=n_ia^{3}$ are, respectively, the entropy and the number of particle in a comoving volume. This means that the condition  $\dot \sigma_i=0$ has a direct physical meaning (in this connection, see also the kinetic covariant approach discussed in \cite{LI2014})
\begin{equation}
\dot S_i/S_i=\dot N_i/N_i = \Gamma  \Rightarrow S_T = \sum_{i=1}^{N} S_i= k_B \sum_{i=1}^{N} N_i.
\end{equation}
Therefore, the entropy growth due to the gravitational particle production process is closely related with the emergence of particles in the space-time thereby leading to the expected enlargement of the  phase space. 

For each fluid component, the energy conservation law ($u_{\mu}T^{\mu \nu}_{(i)};_{\nu}=0$) including particle creation reads:
\begin{equation}\label{ECL}
\dot \rho_i + 3H(\rho_i + p_i + P_{ci})=0,
\end{equation}
a result that also follows directly from EFE  (\ref{EE1})-(\ref{EE2}). It can be integrated once the creation rate is given. In what follows, we assume that at low redshifts the creation pressure is constant, or equivalently (see \cite{LJO2010} for the standard CCDM model, that is, $\Gamma_i=0$ for $i>1$, $\alpha_1=\alpha$ and $\epsilon_i=0$, i=1, 2,..N)  
\begin{equation}\label{CR}
\frac{\Gamma_i}{3H} = {\alpha_i}\frac{\rho_{co}}{\rho_{i}} \Rightarrow  P_{ci} = - (1 + \epsilon_i) \alpha_i \rho_{co}.
\end{equation}
In the above expression, $\alpha_i$ is a dimensionless constant modulating the creation rate of the i-th component which should be determined from quantum field theory in curved spacetimes  and $\rho_{co}$ is the present day value of the critical density. Note also that the constant $P_{ci}$ also depends on the nature of the created components through the kinetic EoS parameter ($\epsilon_i$).


\begin{figure*}\label{fig1}
\centerline{\psfig{figure=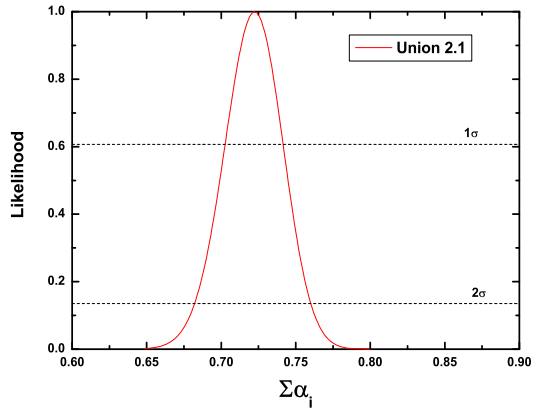,width=3.1truein,height=2.5truein}}
\vspace{-1.1cm}\begin{minipage}[t]{6in} \caption{{\bf{a)}} Likelihood for
$\sum_i{\alpha_{i}}$ based on the
Union2.1 SNe Ia sample. Our statistical analysis performed for a flat matter creation model  
provided $\sum_{i}{\alpha_{i}}= 0.722\pm 0.021$ (1$\sigma$). 
Note that the cosmic history is sensitive only to the net contribution $\sum_{i}{\alpha_{i}}$ whose best fit yields  
 $\Omega^{eff}_m \sim 0.278$ (see expression (14)).}
\end{minipage}
\end{figure*}

Now, by inserting the above expression into the energy conservation law (see Eq. (\ref{ECL})) a direct integration yields:
\begin{equation}\label{CLI}
\rho_{i} = (\rho_{io} - \alpha_i\rho_{co})a^{-3(1 + \epsilon_i)} + \alpha_i\rho_{co}\,, 
\end{equation}
where $\rho_{io}$ is the present day energy density of the {\it i-th} component. The total energy density reads:
\begin{equation}\label{CLI}
\rho_T = \sum_i \rho_{i} = \rho_{co}\left[(\sum_{i=1}^{N}\Omega_{io} - \sum_{i=1}^{N}\alpha_i)a^{-3(1 + \epsilon_i)}+ \sum_{i=1}^{N} \alpha_i\right]\,, 
\end{equation}
{where $\Omega_{io} = {\rho_{io}}/{\rho_{co}}$.}  Now, by combining this result with Eq. (\ref{EE1}) we obtain the Hubble parameter in terms of the redshift z:
\begin{equation}
H^{2} = H_0^{2} \left[(\sum_{i=1}^{N}\Omega_{io} - \sum_{i=1}^{N}\alpha_i)(1+z)^{3(1+\epsilon_i)} + {\sum_{i=1}^{N} \alpha_i} \right]\,,
\end{equation}
where $z\equiv 1/a - 1$. In particular, assuming that all created particles determining the present stage of the Universe are non-relativistic ($\epsilon_i=0$) including even the dark matter component (neither hot nor warm dark matter), {and also using that for a flat Universe  $\sum_i \Omega_{io}=\sum_{i=1}^{N}\rho_{io}/\rho_{co} \equiv 1$,}  the above expression becomes:

\begin{equation}
H^{2} = H_0^{2} \left[{(1 -  \sum_{i=1}^{N} \alpha_i})(1+z)^{3} + {\sum_{i=1}^{N} \alpha_i} \right]\,.
\end{equation}
At this point, it is interesting to compare  the above $H(z)$ expression with the one predicted by the flat concordance model ($\Lambda$CDM), namely:  
\begin{equation}
H^{2}_{\Lambda CDM} = {H_0}^2 \left[(1 - \Omega_{\Lambda})(1+z)^3 + \Omega_{\Lambda}\right]\,, 
\end{equation}
where $\Omega_{\Lambda}$ is the vacuum density parameter and $1 - \Omega_{\Lambda} \equiv \Omega_m = \Omega_{dm} + \Omega_b$ quantifies the  contribution of cold dark matter plus the baryonic components.  One may see that the models have exactly the same Hubble parameter, $H(z)$, just by identifying the multicomponent creation parameter (the hat in ${\hat \Omega}_{\Lambda}$ is to stress its intrinsic creation origin)

\begin{equation}\label{EQUIV}
\sum_{i=1}^{N} \alpha_i = {\hat \Omega}_{\Lambda}\equiv  \Omega_{\Lambda}\,\,\,\, and \,\,\, \Omega^{eff}_{(m)}=1-\sum_{i=1}^{N} \alpha_i\,.   
\end{equation}
As it will be discussed ahead, these are the most important analytical results of the present work. This remarkable dynamic equivalence can also be directly inferred through the differential equation governing the evolution of the scale factor.  By inserting the expression of the creation pressure $P_{ci}$ in the Einstein equation (\ref{EE2}) we obtain:

\begin{equation}
\label{evolR}
2a{\ddot a}+ {\dot{a}}^2 - 3{H_0}^{2}{\sum_{i=1}^{N} \alpha_i} a^{2}  = 0,  
\end{equation} 
which should be compared to the expression 
\begin{equation} \label{evolRLCDM}
2a{\ddot a}+ {\dot{a}}^2 - 3{H_0}^{2}\Omega_{\Lambda}a^{2}  = 0,  
\end{equation}
provided by the $\Lambda$CDM model. Again, the above equations imply that the  same dynamic behavior is recovered when the net contribution of all created components (each one quantified by $\alpha_i$) is identified by the expression previously derived  based on the expression for $H(z)$ (see Eq. (\ref{EQUIV})). 

Nevertheless, since we are working with a multifluid description without dark energy, one may ask about the equivalent matter and vacuum contributions at the level of the constraint Friedman equation (\ref{EE1}). With the help of solution (\ref{CLI}), it is readily checked that the total energy density  behaves like a mixture of  N-components describing non-relativistic matter plus vacuum in the following manner:
\begin{equation}\label{ED}
\rho_T = \rho^{eff}_{(m)}   + \rho^{eff}_{(v)},  
\end{equation}
where $\rho^{eff}_{(m)} = \sum_{i=1}^{N} \rho^{eff}_{(mi)} =  \rho_{co} (1 - \sum_{i=1}^{N}\alpha_i)a^{-3}$ and $\rho^{eff}_v =-P^{eff}_{(v)}= \sum_{i=1}^{N}\alpha_i\rho_{co} \equiv \Lambda^{eff}/8\pi G$ (cf. Eqs. (8) and (10)). Note that the effective vacuum  pressure is just the total negative creation pressure for $\epsilon_i=0$. Only in this case the separation includes naturally a vacuum component whose density parameter was identified before (see relation (14)).

\begin{figure*}\label{fig2}
\centerline {\psfig{figure=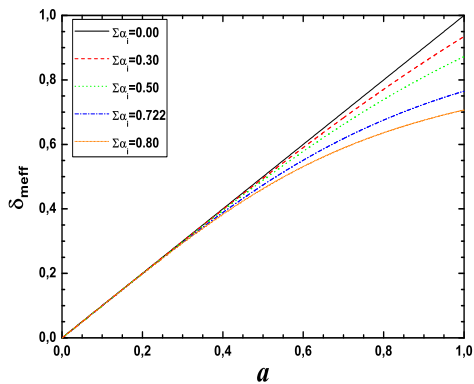,width=3.1truein,height=2.5truein}}
\vspace{-1.4cm}\begin{minipage}[t]{6in} \caption{Evolution of the  density contrast $\delta^{eff}_{(m)}=\delta\rho^{eff}_{(m)}/{\rho^{eff}_{(m)}}$ for different values of the free parameter ($\sum_i{\alpha_{i}}$). The blue line  obtained for the best fit value from SNe Ia data (see Fig. 1), reproduces exactly the  standard $\Lambda$CDM result. Note that  $\delta^{eff}_{(m)}$ involves only that portion of created components which is able to appear as a clustered matter while the effective vacuum medium corresponds to the complementary portion that is  smooth even at the perturbative level (see discussion below (17)). }
\end{minipage}
\end{figure*}
In Figure {\bf 1}, we display the likelihood for the net parameter, 
$\sum{\alpha_{i}}$, based on a flat matter creation Universe driven by N non-relativistic  components. The analysis performed with the SNe Ia  
data (Union2.1 sample \cite{Suzuki2012}) provided the constraint $\sum{\alpha_{i}} = 0.722\pm 0.021$ (1$\sigma$). The
horizontal lines correspond to cuts in the regions of $68.3\%$ and $95.4\%$
probability. For all practical purposes, the present day observed components, namely:  CDM, baryons, neutrinos and CMB photons are nowadays mildly created by the accelerating expanding Universe with different rates quantified by the corresponding creation parameter $\alpha_i$ (i=1,2,3,4). Notice that the cosmic history is sensitive only to the net contribution of all created components, namely, the parameter $\sum_i{\alpha_{i}}$. 

In light of the above constraint, it is also interesting to accentuate the physical role played by the parameter $\sum_i{\alpha_{i}}$ in this wider CCDM framework for an expanding flat geometry ($\Omega_m = 1$). {\it It provides  the effective vacuum energy density} ($\Omega_{\Lambda} \sim 0.7$) {\it and, simultaneously, is also responsible by the suppression on the total matter density parameter so that  its clustered part \it} ($\Omega^{eff}_{(m)} \sim 0.3$) {\it becomes also in agreement with the present observations.} Note also that the formulation works for N-components so that any new (non-relativistic) cosmic component (beyond CDM and baryons) must be accommodated within the constrain derived for the net creation parameter. 


\vspace{0.2cm}

\noindent {\bf 3. Evolution of Perturbations} -  It is widely known that some cosmological models can be in agreement  with the observations at the expanding background 
level although in contradiction with the perturbations data during the linear or nonlinear stages. So, it is mandatory to examine the growth of perturbations for any proposed model before to decide on its physical viability. In the context of the CCDM model such a question was first investigated by Jesus et al.\cite{BL2011} at the linear perturbative level, and later on by Ramos, dos Santos and Waga \cite{Waga2014,Waga2014a} both in the linear and nonlinear stages. 

Based on the so-called neo-Newtonian equations \cite{LZB97}, the first  authors  discussed the evolution of the linear perturbations and compared with the $\Lambda$CDM model prediction. Even being suggested  that the observed density parameter (from galaxy clusters)  should be $\Omega^{eff}_{m}\equiv 1-\alpha$ (see Table 1 and comment in the conclusions  of  Ref. \cite{LJO2010}), a strict one fluid description was assumed in their analysis. By taking  the sound speeds $c_s^{2}=c_{eff}^{2}=0$ so that the neo-Newtonian and relativistic approaches coincide \cite{RRR}, they found that the evolution of the density contrast agree with the one predicted by the $\Lambda$CDM model but only  until $z \gtrsim 1$, after which it is strongly suppressed. 

The second set of authors reanalysed the evolution of perturbations in a more detailed way by using the neo-Newtonian and relativistic approaches for linear evolution\cite{Waga2014}, and the relativistic one for the nonlinear regime\cite{Waga2014a}. By introducing the mentioned separation at the level of fluctuations, it was found that both models (CCDM \& $\Lambda$CDM) have the same skewness signature so that the degeneracy between the CCDM and the $\Lambda$CDM remains at any order in perturbation theory with the proviso that the number of baryons is conserved. 

Here we adopt the same approach, however,  by considering that all non-relativistic components (CDM and baryons) are created. Indeed, the basic results  are not modified even whether an arbitrary number of non-relativistic components are created but only the clustered part is able to agglomerate, that is, only the portion $\rho^{eff}_{(m)}=\rho_{co}(1-\sum_{i} \alpha_i)a^{-3}$ feel the clustering process. In this case, the contribution  of the remaining part, $\rho^{eff}_{(v)} = \sum_{i=1}^{N}\alpha_i\rho_{co}$ represents the smooth vacuum energy density, and, as such, the $\Lambda$CDM equivalence becomes almost trivial. Now, by assuming the above separation and recalling that Newtonian gravity can be applied  for perturbations well inside the horizon, the evolution of the density contrast,  $\delta^{eff}_{(m)}=\delta\rho^{eff}_{(m)}/{\rho^{eff}_{(m)}}$, is driven by the equation: 

\begin{equation}
\ddot\delta^{eff}_{(m)} + 2\frac{\dot a}{a}\dot\delta^{eff}_{(m)} - 4\pi G\rho^{eff}_{(m)}\delta^{eff}_{(m)}=0 \,.
\end{equation}
As one may check, the growing mode solution in terms of the scale factor can be expressed as:
\begin{equation}
\delta^{eff}_{(m)}(a) = C({\bf x})aF\left(\frac13, 1; \frac{11}{6};
-\frac{a^3\sum_i\alpha_i}{1-\sum_i\alpha_i}\right), 
\label{deltam}
\end{equation}
where  $C({\bf x})$ is an integration  time-independent local quantity and $F={}_2F_1(\alpha, \beta, \gamma,z)$  is the Gaussian hypergeometric function. As should be expected, if the net creation parameter $\sum_i\alpha_i\rightarrow 0$ or the scale factor is very small, the results of the standard Einstein-de Sitter model are recovered (${\delta^{eff}_{(m)} \propto a}$, $\Omega^{eff}_{(m)} \rightarrow 1$). 

In Figure {\bf 2}, we show the density contrast $\delta^{eff}_{(m)}$ as a
function of the scale factor for several values of the net creation parameter $\sum_i\alpha_i$. The blue line is associated to the best fit value, $\sum_i{\alpha_{i}}=0.722$,  provided by the SNe Ia observations (see Fig.{\bf 1}). As should be expected it reproduces exactly the prediction of the standard $\Lambda$CDM model for the  density contrast of the non-relativistic matter.  However, it corresponds here to the net density contrast of that portion of the created fluid mixture which is able to grow by gravitational instability. As discussed before,  the smooth created part plays the role of a constant vacuum energy density thereby mimicking the $\Lambda$CDM model both at linear and nonlinear stages (in this connection see also \cite{Waga2014a} for a different but related approach).     

At this point, one may ask on the possibility of breaking the model degeneracy with $\Lambda$CDM. More precisely: {\it Can the existing or future cosmological probes provide a crucial test confronting $\Lambda$CDM and extended CCDM?} Our results suggest that tests involving only the cosmic history or matter perturbations are in principle discarded. However, since the temperature law may be affected by adiabatic creation of photons,  one may expect to break the degeneracy based on probes related to the thermodynamic sector, among them: (i) the CMB thermodynamics and angular power spectrum (ii) Sunyaev-Zeldovich effect, and (iii) the evolution equation of thermal relics (in this connection see Refs.\cite{LI2014,LS2000,Nort2011}).    

\vspace{0.1cm}

\noindent {{\bf 4. Conclusions} - In this paper we have extended the relativistic CCDM cosmology \cite{LJO2010} by including gravitationally induced creation of all possible species existing in the observed Universe. Only non-relativistic components (baryons + CDM) are now responsible by the accelerating cosmic expansion. Since the overall dynamics is fully degenerated with $\Lambda$CDM (at background and perturbative levels), this extended CCDM cosmology suggests that the flat $\Lambda$CDM model is an effective cosmology. Potential probes breaking the model degeneracy were also pointed out and it will be discussed elsewhere. Let us now summarize the main results derived here.  

(1) The previous CCDM paper \cite{LJO2010} (and results based on it) is a particular case where only creation of cold dark matter is taken into account. It is readily recovered by taking  $N=2$, $\alpha_1 \neq 0$) and $\alpha_2 = 0$ (baryons are not created), and  $\sum_{i} \alpha_i = \alpha_1$ ($\alpha$, in their notation) playing the dynamical role of $\Omega_{\Lambda}$. 


(2) All created components are treated in the same manner in agreement with the universality of the gravitational interaction (equivalence principle). We stress that an unknown number of non-relativistic components (including baryons and CDM) can be accommodated within the observational limit of the net free parameter $\sum_{i=1}^{N}\alpha_i$ (see Figs. 1, 2). All possibilities are counted by the effective parameter $\sum_i \alpha_i$ which provides the present net creation $[\sum_i\Gamma_i\rho_i]_{today} \sim [\sum_i\alpha_i\rho_{co}H_0] \sim  10^{-39}g.cm^{-3}.yr^{-1}$.  {Such very tiny homogeneously distributed  creation rate in the Universe  ($\sim 10^{-15}$ nucleon/$cm^{3}.yr$) is not locally accessible by present day experiments.}

Finally, we remark that the dynamical equivalence of this extended CCDM  scenario with $\Lambda$CDM  suggests that the observed Universe is somehow selecting the production of non-relativistic particles. In light of the above results, one may think that instead to be erased by dark energy, all the non-relativistic matter content is being continuously replenished by the late time expanding Universe.
  
{\bf Acknowledgments} The authors are also grateful to Ioav Waga, Iuri Baranov and Supriya Pan for helpful discussions. JASL, RCS and JVC are partially supported by FAPESP, CNPq and CAPES (LLAMA project, INCT-A and PROCAD2013 projects).

\end{document}